# Photoinduced suppression of the ferroelectric instability in PbTe

M.P. Jiang[1,2,3], M. Trigo[1,2], S. Fahy[4], É.D. Murray[4], I. Savić[4], C. Bray[1,5], J. Clark[1], T. Henighan[1,2,3], M. Kozina[1,2,5], M. Chollet[5], J.M. Glownia[5], M. Hoffmann[5], D. Zhu[5], O. Delaire[7], A.F. May[7], B.C. Sales[7], A.M. Lindenberg[1,2,8], P. Zalden[1,2,8], T. Sato[9], R. Merlin[10], and D.A. Reis[1,2,5].

[1] Stanford PULSE Institute, SLAC National Accelerator Laboratory, Menlo Park, CA 94025, USA
[2] Stanford Institute for Materials and Energy Sciences, SLAC National Accelerator Laboratory, Menlo Park, CA 94025, USA
[3] Department of Physics, Stanford University, Stanford, CA 94305, USA
[4] Tyndall National Institute and Department of Physics, University College, Cork, Ireland
[5] Department of Applied Physics, Stanford University, Stanford, CA 94305, USA
[6] Linac Coherent Light Source, SLAC National Accelerator Laboratory, Menlo Park, CA 94025, USA
[7] Materials Science and Technology Division, Oak Ridge National Laboratory, Oak Ridge, Tennessee 37831, USA
[8] Department of Materials Science and Engineering, Stanford University, Stanford, CA 94305, USA
[9] School of Science, The University of Tokyo, 7-3-1 Hongo, Bunkyo-ku, Tokyo 113-0033, Japan
[10] Department of Physics, University of Michigan, Ann Arbor, Michigan 48109, USA

**The interactions between electrons and phonons drive a large array of technologically relevant material properties including ferroelectricity, thermoelectricity, and phase-change behaviour. In the case of many group IV-VI, V, and related materials, these interactions are strong and the materials exist near electronic and structural phase transitions. Their close proximity to phase instability produces a fragile balance among the various properties[1-4]. The prototypical example is PbTe whose incipient ferroelectric behaviour[5,6] has been associated with large phonon anharmonicity and thermoelectricity. Experimental measurements[7-10] on PbTe reveal anomalous lattice dynamics, especially in the soft transverse optical phonon branch. This has been interpreted in terms of both giant anharmonicity[7,11-14] and local symmetry breaking due to off-centering of the Pb ions[8,9]. The observed anomalies have prompted renewed theoretical and computational interest[11-20], which has in turn revived focus on the extent that electron – phonon interactions drive lattice instabilities in PbTe and related materials. Here, we use Fourier-transform inelastic x-ray scattering (FT-IXS)[21,22] to show that photo-injection of free carriers stabilizes the paraelectric state. With support from constrained density functional theory (CDFT) calculations[23], we find that photoexcitation weakens the long-range forces along the cubic direction tied to resonant bonding and incipient ferroelectricity[20]. This demonstrates the importance of electronic states near the band edges in determining the equilibrium structure[2-4,24].**

The incipient ferroelectric behavior of the IV-VI compounds is strongly sensitive to carrier concentration[25]. This implies that the coupling of electronic states near the band edge with the soft phonon branch plays a defining role in the equilibrium structure. In order to gain further insight into these interactions, we use ultrafast infrared excitation across the direct band-gap to transiently affect the carrier concentration at fixed lattice spacing. This modifies the band-population that in turn affects the phonon frequencies, while mixing the transverse acoustic and optical branches. We measure the lattice vibrational response of the transient photo-excited state through Fourier-transform inelastic X-ray scattering (FT-IXS)[21,22]. FT-IXS is a new time and momentum resolved pump-probe based method for obtaining lattice dynamics. Here absorption of the pump pulse produces correlated pairs of phonons with equal and opposite momenta that are probed by femtosecond x-ray diffuse scattering. The femtosecond x-ray diffuse scattering intensity is proportional to snapshots of these phonon-phonon correlations as a function of momentum transfer. The amplitude of the oscillatory component is directly related to the strength of the interaction of the phonons with the photo-excited electron density and thus provides unique information about the nature of the ferroelectric instability.

The experiment was performed at the XPP instrument at the Linac Coherent Light Source (LCLS)[26] free-electron laser. We excite ~2 x $10^{20}$ cm$^{-3}$ carriers per L-valley (~0.5% valence excitation) across the direct band gap of n-type PbTe with 60 fs infrared (IR) pump pulses with central wavelength of 2 μm (see Methods). Subsequently, we track the x-ray diffuse scattering as a function of time delay τ between the IR pump and x-ray probe pulses. Figure 1 shows the time-domain data extracted from relative differential changes in the x-ray diffuse scattering intensities, $\delta I(\tau;\mathbf{Q})/I_0(\mathbf{Q})$ along approximately the Δ direction (Γ to X), or $\mathbf{q} = (0 -q_y 0)$. Here $I_0(\mathbf{Q})$ represents the diffuse scattering at momentum transfer $\mathbf{Q = G+q}$ when the pump arrives after the x-ray probe. From top to bottom the data in Fig. 1a span from near zone-center to zone edge in the $\mathbf{G}$ = (-1 1 3) Brillouin zone (BZ) and in Fig. 1c from near zone center to approximately half the distance to zone edge in the $\mathbf{G}$ = (-2 0 4) BZ. All momenta/wavevectors are given in reciprocal lattice units (rlu) for the conventional face-centered cubic Bravais lattice (and thus conventional body-centered cubic reciprocal lattice). In our experimental geometry, the diffuse scattering along Δ is sensitive primarily to transverse phonons polarized along (0 0 1). The data show coherent, damped oscillatory behavior consisting of multiple frequencies that disperse with increasing reduced wavevector $\mathbf{q}$ on top of a picosecond-scale overall decay.

In Figures 1b and 1d, we plot the Fourier transform of the time-domain data for the (-1 1 3) and (-2 0 4) zones as a function of $|q_y|$ (on a logarithmic color scale). Several dispersive features are immediately apparent in the spectra on top of a broadband and relatively featureless signal originating from the overall decay of differential diffuse scattering over time. The dispersive features in Figure 1b match the first transverse acoustic overtone (2TA) and transverse optical and acoustic combination modes (TO±TA), extracted from the INS data in Cochran[27] (overlaid in red). The overtone is due to phonon-phonon correlations between modes at +q and –q within the same branch[21], while the combination modes originate from correlations between different branches. These mixed terms were not observed in previous FT-IXS measurements [ref. 21, 22] but are expected in general where photoexcitation differentially changes various elements of the dynamical matrix. The absence of a signal at 2TO is likely due to its expected short lifetime (half of the already short TO) leading to a broad frequency signal that is below our detection limit. The predominance of the combination modes indicates that photoexcitation couples the TO and TA eigenvectors in the nonequilibrium state and does not just renormalize the phonon frequencies, as would results from a simple scaling of the dynamical matrix. This is consistent with a picture where photoexcitation primarily affects the long-range interactions that have been linked to the ferroelectric instability through resonant bonding[20]. As shown below, our constrained density functional theory calculations support this picture.

In the (-2 0 4) BZ, the sensitivity to optical phonons is low and we therefore only expect to see overtones of the acoustic modes, consistent with the results shown in Figure 1d. Here, we attribute the dispersion to 2TA. The deviation from the neutron results (red line) is likely due to small deviations from the Δ-line for this zone in our geometry (see Methods).

To simulate the effects of the pump pulse, we perform constrained density functional theory (CDFT) calculations in which 1% of valence electrons are suddenly promoted into the conduction band (see Methods). In Figure 2a, we juxtapose the results for the excited state dispersion from our CDFT calculations for the overtone and combination modes along Δ in the (-1 1 3) BZ (orange traces) with the background-subtracted FT-IXS data from Fig. 1b. The dispersion of the electronically unexcited state is obtained from DFT (red traces) is also displayed for reference. The experimental data qualitatively matches the CDFT traces, reflecting the closeness of the model in depicting the effects of ultrafast photoexcitation on the material. To further emphasize the correspondence, we use the excited state force constants to derive the resultant dynamics of the phonon squared displacements due to a sudden photoexcitation of carriers, and extract the FT-IXS spectrum as shown in Figure 2b (on the same color scale). As with Fig. 2a, we overlay the overtone and combination mode dispersions from CDFT and DFT for comparison. It is clear from comparing Figs. 2a and 2b that the simulated spectrum captures the general features of the experimental results closely.

Figure 2c shows a time-domain comparison between experimental (black lines) and calculated (orange lines) dynamics at select coordinates along Δ in the (-1 1 3) BZ. Note that the model calculations do not account for relaxation of the photoexcited carriers or the heating of the lattice, which could explain the picosecond rise seen in the data but not the model. Importantly, the strong initial dip and phase of the oscillation in the experimental intensity is well reproduced by the model. In the calculations, this coincides with substantial hardening of the TO phonon branch of up to ~48% at Γ, driven by a distinctive weakening of long-range interactions involving both 4th and 8th nearest neighbors as illustrated in Figure 3[1]. Here, Fig. 3a shows the calculated ground state and photoexcited harmonic force constants as blue dots and red crosses respectively. The differential force constants for each nearest neighbor interaction are also conveniently shown in Fig. 3b. Photoexcitation notably reduces the absolute magnitudes of the 4th and 8th nearest neighbor force constants. Equivalently, these particular force constants correspond to interactions with the 2nd and 3rd neighbors along the <100> direction in the cubic crystal structure of PbTe. The calculations reproduce the anomalously strong long-range interactions that are expected along this direction, and at-tributed to resonant bonding involving *p*-band valence electrons[20]. It is further observed that the experimental lifetimes are comparable, but perhaps shorter than those predicted in the simulations based on CDFT calculations, where we used the 3rd order anharmonic interactions for the ground state.

---

[1] The appearance of a sizable interaction with the 8th nearest neighbor can be observed in computations on 216-atom supercells. This differs from the calculations displayed in Figure 2, which are only on 64-atom supercells. However, the phonon dispersions derived from either supercell size show no prominent differences.

The identifications of the two-phonon dispersion for the transient state combined with the CDFT calculations indicate that photoexcitation of PbTe stabilizes the paraelectric phase. The specific observation of all-transverse combination modes along the Δ direction is consistent with differential changes in the real-space force constants, as obtained in the CDFT calculations. Furthermore, the initial rise of the oscillation after τ = 0 in the (-2 0 4) BZ (Fig. 1c) is consistent with a sudden softening of the TA branch frequency upon photoexcitation. This conclusion can be drawn independently from the calculations (which show the same effect, see Fig. 2a) since the square of the phonon displacements, and thus the diffuse scattering, is inversely proportional to the phonon frequency[21]. In contrast, the initial dip of the response in the (-1 1 3) BZ (Fig. 1a) likely originates from a hardening of the TO branch after photoexcitation, as predicted by CDFT calculations (see Fig. 2a). We conclude that the impulsive redistribution of the electronic states near the Fermi energy is consistent with a weakening of the long-range interactions along the <100> cubic direction found in the CDFT calculations. This stems from photo-excitation of delocalized carriers from p-like resonant bonding states to anti-bonding states, thus reducing the propensity for a ferroelectric distortion[24]. This immediately softens the TA branch while hardening the TO branch near zone center. These frequency shifts arise in the CDFT calculation of harmonic force constants, independent of anharmonic effects, indicating that the soft mode behavior is a consequence of the strong electron-phonon coupling.

Our results demonstrate the considerable role that electronic states near the Fermi level play in the lattice dynamics of PbTe. The experimental data is consistent with a strong hardening of the soft TO mode frequency following ultrafast photoexcitation that disrupts the resonant bonds in the material. Essentially, this moves PbTe further from its incipient ferroelectric state, stabilizing the paraelectric phase. Based on this, the soft character of the TO phonon can evidently be linked to its coupling with the electronic states near the band edges in the system.

1. Kawamura, H. Phase transition in IV-VI compounds. *Narrow Gap Semiconductors Phys. and App. Lecture Notes in Physics Volume* **133**, 470-494 (1980).
2. Cohen, M.H., Falicov, L.M. and Golin, S. Crystal chemistry and band structures of the group V semimetals and the IV-VI semiconductors. *IBM Journal of Research and Development* **8(3)**, 215-227 (1964).
3. Littlewood, P.B. The crystal structure of IV-VI compounds: I. Classification and description. *J. Phys. C: Solid St. Phys.* **13**, 4855-4873 (1980).
4. Littlewood, P.B. The crystal structure of IV-VI compounds: II. A microscopic model for cubic/rhombohedral materials. *J. Phys. C: Solid St. Phys.* **13**, 4875-482 (1980).
5. Alperin, H.A., Pickart, S.J., Rhyne, J.J. & Minkiewicz, V.J. Softening of the transverse-optic mode in PbTe. *Phys. Lett. A* **40**, 295-296 (1972).
6. Bate, R.T., Carter, D.L. & Wrobel, J.S. Paraelectric behavior of PbTe. *Phys. Rev. Lett.* **25**, 159-162 (1970).
7. Delaire, O. *et al.* Giant anharmonic phonon scattering in PbTe. *Nature Materials* **10**, 614-619 (2011).
8. Bozin, Emil S. *et al.* Entropically stabilized local dipole formation in lead chalcogenides. *Science* **330**, 1660-1663 (2010).
9. Jensen, Kirsten M.O. *et al.* Lattice dynamics reveals a local symmetry breaking in the emergent dipole phase of PbTe. *Phys. Rev. B* **86**, 085313 (2012).
10. Burkhard, H., Bauer, G. & Lopez-Otero, A. Submillimeter spectroscopy of TO-phonon mode softening in PbTe. *J. Opt. Soc. Am.* **67 (7)**, 943-946 (1977).
11. Shiga, T. *et al.* Microscopic mechanism of low thermal conductivity in lead telluride. *Phys. Rev. B* **85**, 155203 (2012).
12. Shiga, T., Murakami, T., Hori, T., Delaire, O. & Shiomi, J. Origin of anomalous anharmonic lattice dynamics of lead telluride. *Appl. Phys. Express* **7**, 041801 (2014).
13. Chen, Y., Ai, X. & Marianetti, C.A. First-principles approach to nonlinear lattice dynamics: Anomalous spectra in PbTe. *Phys. Rev. Lett.* **113**, 105501 (2014).
14. Li, C.W. *et al.* Phonon self-energy and origin of anomalous neutron scattering spectra in SnTe and PbTe thermoelectrics. *Phys. Rev. Lett.* **112**, 175501 (2014).
15. Zhang, Y., Ke, X., Kent, P. R.C., Yang, J. & Chen, C. Anomalous lattice dynamics near the ferroelectric instability in PbTe. *Phys. Rev. Lett.* **107**, 175503 (2011).
16. Ekuma, C.E., Singh, D.J., Moreno, J. & Jarrell, M. Optical properties of PbTe and PbSe. *Phys. Rev. B* **85**, 085205 (2012).
17. Tian, Z. *et al.* Phonon conduction in PbSe, PbTe, and PbTe$_{1-x}$Se$_x$ from first-principles calculations. *Phys. Rev. B* **85**, 184303 (2012).
18. Nielsen, M.D., Ozolins, V. & Heremans, J.P. Lone pair electrons minimize lattice thermal conductivity. *Energy Environ. Sci.* **6**, 570-578 (2013).
19. Alves, H.W.L., Neto, A.R.R., Scolfaro, L.M.R., Myers, T.H. & Borges, P.D. Lattice contribution to the high dielectric constant of PbTe. *Phys. Rev. B* **87**, 115204 (2013).
20. Lee, S. *et al.* Resonant bonding leads to low lattice thermal conductivity. *Nature Comm.* **5**, 3525 (2014).
21. Trigo, M. *et al.* Fourier-transform inelastic X-ray scattering from time- and momentum-dependent pho-

non-phonon correlations. *Nature Phys.* **9**, 790-794 (2013).
22. Zhu, D. *et al.* Phonon spectroscopy with sub-meV resolution by femtosecond x-ray diffuse scattering. *Phys. Rev. B* **92**, 054303 (2015).
23. Murray, É.D., Fritz, D.M., Wahlstrand, J.K., Fahy, S. & Reis, D.A. Effect of phonon anharmonicity in photoexcited bismuth. *Phys. Rev. B* **72**, 060301 (2005).
24. Littlewood, P.B. and Heine, V. The infrared effective charge in IV-VI compounds: I. A simple one-dimensional model. *J. Phys. C: Solid State Phys.* **12**, 4431-4439 (1979).
25. Jantsch, W. Electronic and Dynamical Properties of IV-VI Compounds. In *Springer Tracts in Modern Physics Vol. 99* (Springer-Verlag, Berlin, 1983).
26. Chollet, M. *et al.* The X-ray Pump-Probe instrument at the Linac Coherent Light Source. *J. Sync. Rad.* **22**(3), 503-507 (2015).
27. Cochran, W., Cowley, R.A., Dolling, G. & Elcombe, M.M. The crystal dynamics of lead telluride. *Proc. R. Soc. Lond. A* **293**, 433-451 (1966).

**Acknowledgments.** This work is supported by the Department of Energy, Office of Science, Basic Energy Sciences, Materials Sciences and Engineering Division, under Contract DE-AC02-76SF00515. S. Fahy, É.D. Murray, and I. Savić acknowledge support from ***. O. Delaire acknowledges support from the U.S. Department of Energy, Office of Science, Basic Energy Sciences, Materials Science and Engineering Division, through the Office of Science Early Career Research Program. A.F. May and B.C. Sales were supported by the U.S. Department of Energy, Office of Science, Basic Energy Sciences, Materials Science and Engineering Division. J.C. acknowledges financial support from the Volkswagen Foundation. Portions of this research were carried out at the Linac Coherent Light Source (LCLS) at the SLAC National Accelerator Laboratory. LCLS is an Office of Science User Facility operated for the U.S. Department of Energy Office of Science by Stanford University. Preliminary experiments were performed at SACLA with the approval of the Japan Synchrotron Radiation Research Institute (JASRI) (Proposal No. 2013A8038) and at the Stanford Synchrotron Radiation Lightsource, SLAC National Accelerator Laboratory, which like the LCLS is supported by the U.S. Department of Energy, Office of Science, Office of Basic Energy Sciences under Contract No. DE AC02-76SF00515.

**Author Contributions.** D.A.R. conceived this project. M.P.J. and M.T. led the experiment, supported by C.B., M.C., J.C., O.D., J.M.G., T.H., M.H., M.K., A.M.L., P.Z., D.Z., and D.A.R. I.S, É.D.M. and S.F., performed the DFT and CDFT calculations and contributed to the interpretation along with D.A.R., M.P.J., M.T., O.D. and R.M. The sample was grown and characterized by O.D., A.F.M., and B.C.S. T.S. helped with preliminary x-ray experiments. The manuscript was written by M.P.J., M.T., and D.A.R., with feedback from all co-authors.

**Author Information.** Reprints and permissions information is available at www.nature.com/reprints. The authors declare no competing financial interests. Readers are welcome to comment on the online version of the paper. Correspondence and requests for materials should be addressed to D.A.R. (dreis@slac.stanford.edu) or M.P.J. (mpjiang@stanford.edu).

## METHODS

**Experimental details.** The reported measurements were performed at the X-ray pump probe (XPP) instrument of the Linac Coherent Light Source (LCLS) X-ray free-electron laser (FEL) using infrared pump pulses (60 fs, 350 µJ, 0.6 eV) generated from an optical parametric amplifier laser and x-ray probe pulses (50 fs, 8.7 keV) with energies selected using a diamond (111) monochromator[28] leading to a bandwidth of ~0.5 eV. Both beams approached the sample at a grazing incidence angle (< 5°) and the entire sample space was contained within a He-filled chamber to minimize background air scattering. Preliminary measurements were also taken at the Stanford Synchrotron Radiation Lightsource (SSRL) (X-ray only) and the Spring-8 Angstrom Compact Free Electron Laser (SACLA) (time-resolved, 1.5 eV optical pump / 9 keV x-ray probe). A large area, 2.3 Mpixel Cornell-SLAC Pixel Array Detector (CSPAD) with $110 \times 110$ µm$^2$ pixels and a 120 Hz readout rate captured the resulting x-ray diffuse scattering.

Diffuse scattering patterns were recorded at a repetition rate of 120 Hz and filtered on a shot-by-shot basis according to fluctuations in both the electron beam energy and x-ray intensity. Images passing the filter constraints were subsequently sorted into 100 fs-wide time delay bins according to the per-shot output of a timing diagnostic tool accurately tracking the relative arrival time between pump and probe[29]. The resulting images in each bin are summed and normalized by the sum of x-ray intensities of the shots corresponding to those images. The reported results stem from the analysis of an average of ~10,000 shots per time delay bin, which accounting for unused, filtered-out shots, overhead in mechanical movement speeds, and FEL machine downtime, is collected in a matter of several hours in real-time.

Sets of room temperature (RT) measurements were taken at two infrared pump fluences, differing by a factor of approximately two. Considering uncertainties in the incidence angle and size of the focused beam, the estimated photoexcited carrier density of the higher pump fluence (the reported data) is $2 \times 10^{20}$ cm$^{-3}$. Other than differences in the amplitude of the differential intensity time-domain responses linear with the pump fluence, no notable variations are seen in a comparison of data from the two fluences. Furthermore, the damage limit of the PbTe samples, from exposure to the combined fluence of both infrared and x-ray laser beams, was assessed at both SACLA and LCLS. The chosen aggregate pump and probe fluence for the measurements was thus kept below this limit. No evidence of sample damage was observed during the experiment.

**Sample growth.** Single crystals of PbTe were grown by a modified Bridgman technique with {100} crystallographic orientation at Oak Ridge National Laboratory with n-type carrier concentration of 4 x $10^{17}$ cm$^{-3}$.

**Geometry and data extraction.** We tuned the geometric configuration of the experimental setup to simultaneously capture the approximate projection of the high-symmetry Δ (Γ to X) wavevector direction on the detector in both an all-even (-2 0 4) and an all-odd (-1 1 3) Brillouin zone (BZ), with select sensitivity to transverse phonon polarization scattering contributions. Moreover, the approximate zone center (Γ) peak of the (-1 1 3) BZ was captured by the detector in this configuration, adjusted 1% away from the exact scattering conditions to prevent damage to the detector.

We evaluate diffuse scattering intensity dynamics along reduced wavevectors away from zone center (towards zone edge). In order to characterize the activity away from zone center with sufficient signal-to-noise contrast, multiple neighboring pixels are binned and averaged together to represent specific coordinates along the subsequent reduced wavevector directions. The Δ direction of the (-1 1 3) BZ navigates from q ~ [0 -0.1 0] to q ~ [0 -1 0] in [0 -0.05 0] steps, with each coordinate represented by the closest (in reciprocal space) 50 neighboring pixels, while the Δ direction of the (-2 0 4) BZ moves from q ~ [0 -0.05 0] to q ~ [0 -0.425 0] in [0 -0.025 0] steps, with each coordinate represented by the closest 10 neighboring pixels. It should be noted that while the vast majority of the distribution of $q_x$ and $q_z$ coordinate values along the Δ direction of the (-1 1 3) BZ remain close to zero throughout the entire reduced wavevector, the same coordinate values diverge slightly traversing the Δ direction of the (-2 0 4) BZ. This is due to how the detector cuts the BZ in projecting it into a two-dimensional plane. The deviation from zero for $q_x$ is ~ -0.003 rlu for each q coordinate (segment of binned pixels) moving away from Γ and for $q_z$ is ~ -0.002 rlu. This divergence explains the mismatch with the overlaid Δ direction INS data in Figure 1d.

To produce the FT-IXS dispersion plots depicting the data simultaneously in frequency and momentum space in Figures 1b and 1d, we perform a Fourier transform (FT)-based analysis to translate the time-domain data into the frequency-domain. Specifically, we apply a Fast Fourier transform (FFT) algorithm to the time-domain data extracted at each q coordinate of a reduced wavevector under investigation. Prior to employing the algorithm, additional points are padded to each time-domain trace, effectively extending the time delay range to artificially increase the frequency-domain resolution for visual clarity. The padded points in each trace take on the value of the last recorded point in the experimental data. For the reported data, 175 padded points are used for the away from zone center Δ direction traces. The resulting magnitudes of the outputted complex Fourier coefficients are illustrated in false coloring in both momentum and frequency space on a base-10 logarithmic scale (FT-IXS dispersion figures).

**Constrained DFT calculations.** The harmonic (2$^{nd}$ order) and anharmonic (3$^{rd}$ order) force constants were calculated using the constrained density functional theory (CDFT) approach described in Murray *et al*[23] and DFT calculations respectively. The absorbed photon energy was assumed to be 1.5 eV and carriers were assumed to (separately) thermalize within the valence and conduction bands, keeping the total carrier density in the conduction band fixed at 1% of the valence electrons (1.5 x $10^{21}$ cm$^{-3}$). Thus, carrier recombination between conduction and valence bands was assumed to be negligible in the timescale of the experiments. The 2$^{nd}$ and 3$^{rd}$ order force constants were calculated from atomic forces using a real-space finite difference supercell approach[30] and the Phono3py code[31]. All CDFT ad DFT calculations were carried out using the Abinit code[32] and employing the local density approximation and Hartwigsen-Goedecker-Hutter norm-conserving pseudopotentials. Forces were computed on 64 (216) atom supercells, using an energy cut-off of 15 Ha and 4-shifted 4x4x4 (2x2x2) reciprocal space grids for electronic states. Phonon frequencies, mode eigenvectors and phonon decay rates were calculated for reduced wavevectors throughout the Brillouin zoneas described in He *et al*[33].

The x-ray diffuse scattering signal arising from pure phonon squeezing for reduced wave vectors **q** along symmetry lines in the Brillouin zone was calculated by taking the overlap $u_{q,\lambda}^{\dagger} \Delta D^{-1} u_{-q,\lambda'}$ of the change, $\Delta D^{-1} = D_{after}^{-1}(q) - D_{before}^{-1}(q)$, in the inverse dynamical matrix before and after photoexcitation with mode eigenvectors, $u_{q,\lambda}$ and $u_{-q,\lambda'}$ of the photoexcited system. This determines the cross-correlation between modes λ and λ' immediately following photoexcitation, which then gives rise to damped oscillations at the sum and difference frequencies, $\omega_{q,\lambda} \pm \omega_{-q,\lambda'}$, the damping rate being the average of the damping rates of the two modes involved. These damped oscillations of the cross-correlation contribute to the x-ray diffuse scattering signal for momentum transfer, $\Delta k = q + G$, scaled by the polarization factor $\Delta k \cdot u_{q,\lambda}$ of each mode. Details of this method are in Fahy *et al.*[34].

**References.**

28. Zhu, D. *et al*. Performance of a beam-multiplexing diamond crystal monochromator at the Linac Coherent Light Source. *Rev. Sci Instr.* **85**(6), 063106 (2014).

29. Harmand, M. *et al*. Achieving few-femtosecond time-sorting at hard X-ray free-electron lasers. *Nature Photonics* **7**, 215-218 (2013).

30. Esfarjani, K., Chen, G. & Stokes, H.T. Heat transport in silicon from first-principles calculations. *Phys. Rev. B* **84**, 085204 (2011).

31. Chaput, L., Togo, A., Tanaka, I. & Hug, G. Phonon-phonon interactions in transition metals. *Phys. Rev. B* **84**, 094302 (2011).

32. Gonze, X. *et. al.* ABINIT: First-principles approach to material and nanosystem properties. *Comput. Phys. Comm.* **180**, 2582-2615 (2009).

33. He, Y., Savić, I., Donadio, D. & Galli, G. Lattice thermal conductivity of semiconducting bulk materials: atomistic simulations. *Phys. Chem. Chem. Phys.* **14**, 16209-16222 (2012).

34. Fahy, S., Murray, É.D. & Reis, D.A. submitted for publication to *Phys. Rev. Lett.* (2015).

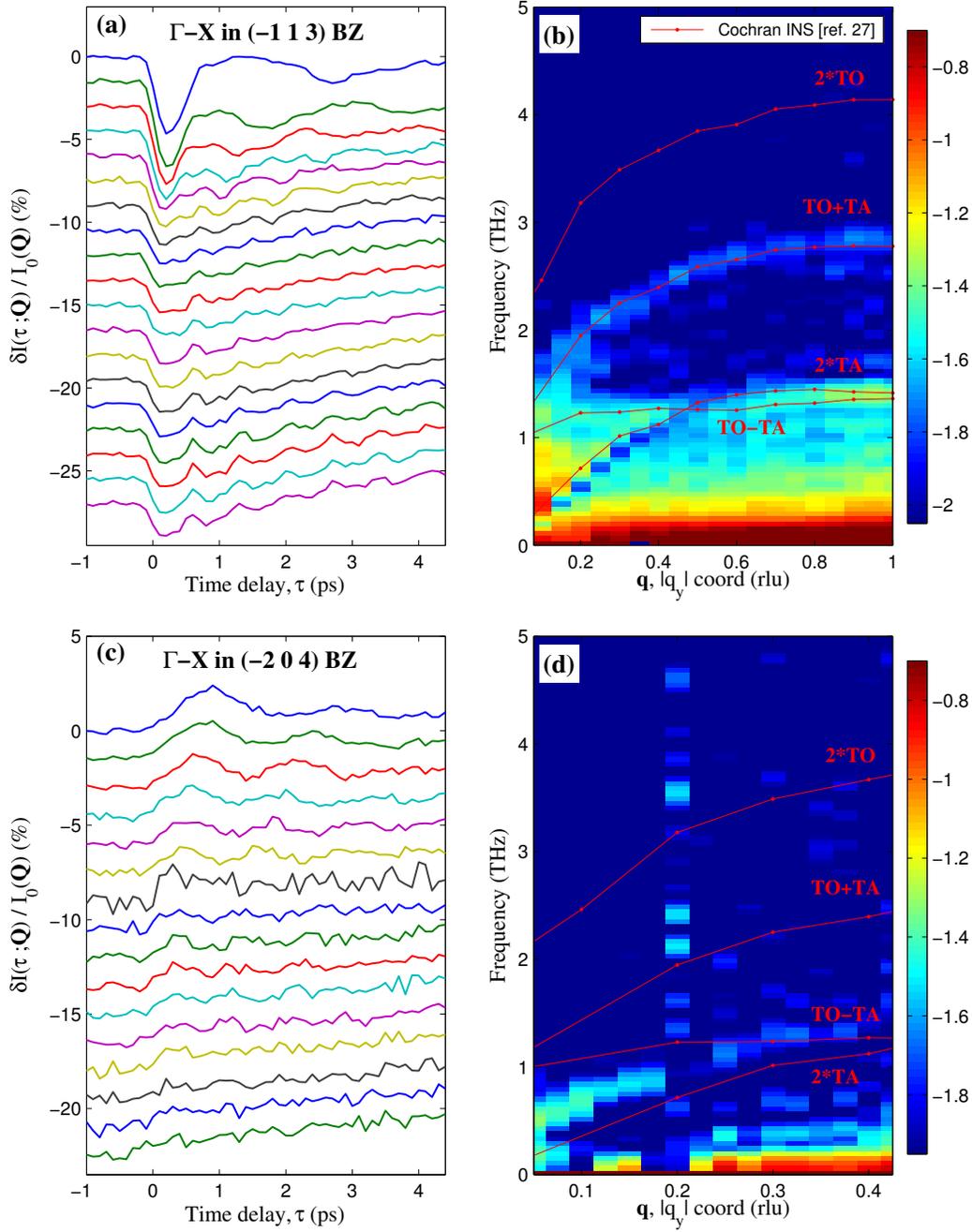

**Figure 1 | FT-IXS spectra away from zone-center in both an all-odd and an all-even Brillouin zone. a,c,** Time-domain traces tracking relative differential x-ray diffuse scattering intensities at various q coordinates along the approximate Γ to X (Δ) high-symmetry reduced wavevector direction in both the (-1 1 3) Brillouin zone (BZ) and the (-2 0 4) BZ respectively. Changes in the x-ray diffuse scattering intensity are instigated by an ultrafast infrared pump pulse of central wavelength 2 μm and the observed modulations stem from temporal coherences in the momentum-dependent phonon-phonon correlations. **b,d,** Simultaneous frequency and momentum representations of the time-domain data in Figs. 1a and 1c respectively, with extracted INS data from ref. 27 overlaid. Clear matches with first overtone and combination transverse polarized phonon states are seen. Note that broadband behavior seen near $|q_y|\sim0.2$ in panel **d** is due to noisier statistics that arise from an average over a smaller solid angle, in comparison with data at other q coordinates.

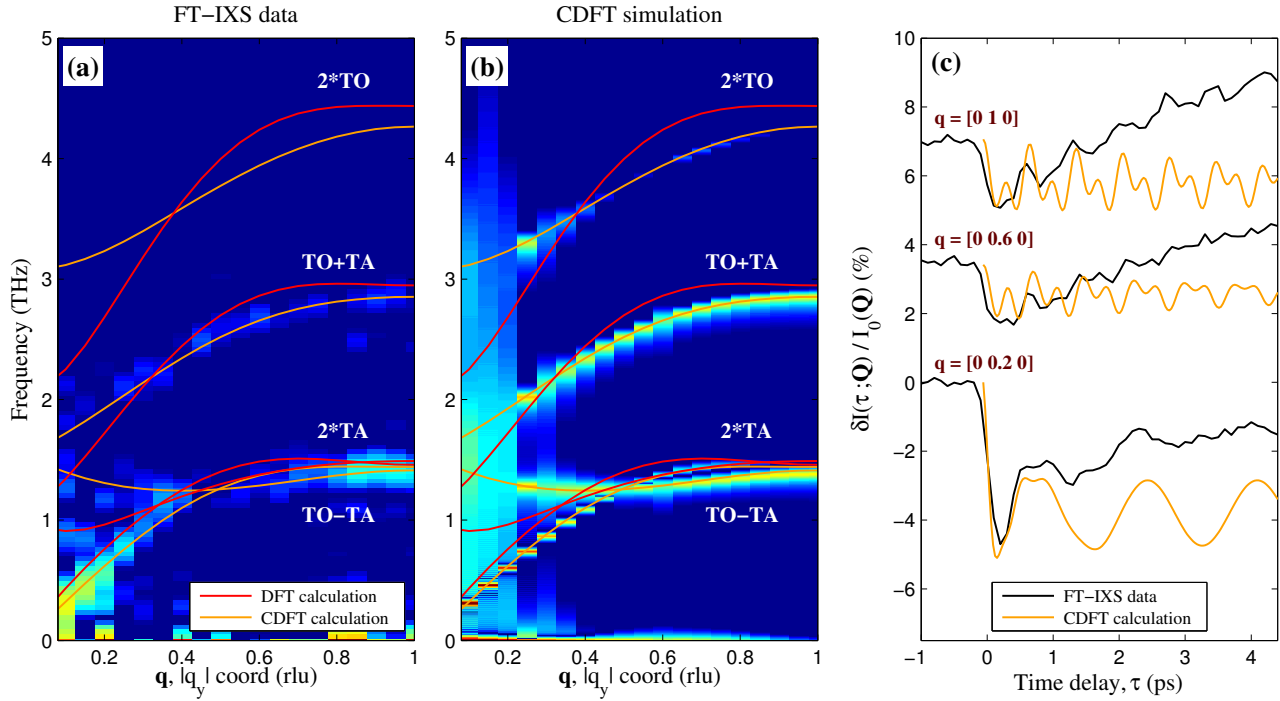

**Figure 2 | Comparison of (-1 1 3) BZ Δ direction results with constrained density functional theory (CDFT) calculations. a,** Background-subtracted version of the FT-IXS spectrum along the Δ direction in the (-1 1 3) BZ seen in Fig. 1b with both CDFT (orange traces) and DFT (red traces) calculations overlaid for comparison. Satisfactory matches are found with the CDFT overtone and combination mode calculations. **b,** Simulated FT-IXS spectrum along the same high-symmetry direction in the (-1 1 3) BZ using CDFT. The CDFT and DFT traces are again overlaid for comparison. The spectrum qualitatively matches the experimental data in Fig. 2a. **c,** Comparison of experimental time-domain traces (black traces) at select wavevector coordinates along Δ as taken from Fig. 1a with traces calculated via CDFT (orange traces). There is a good qualitative match between experimental and model results, notably in the strong initial dip seen at $\tau = 0$, connected to a considerable hardening of the TO branch frequencies in CDFT calculations.

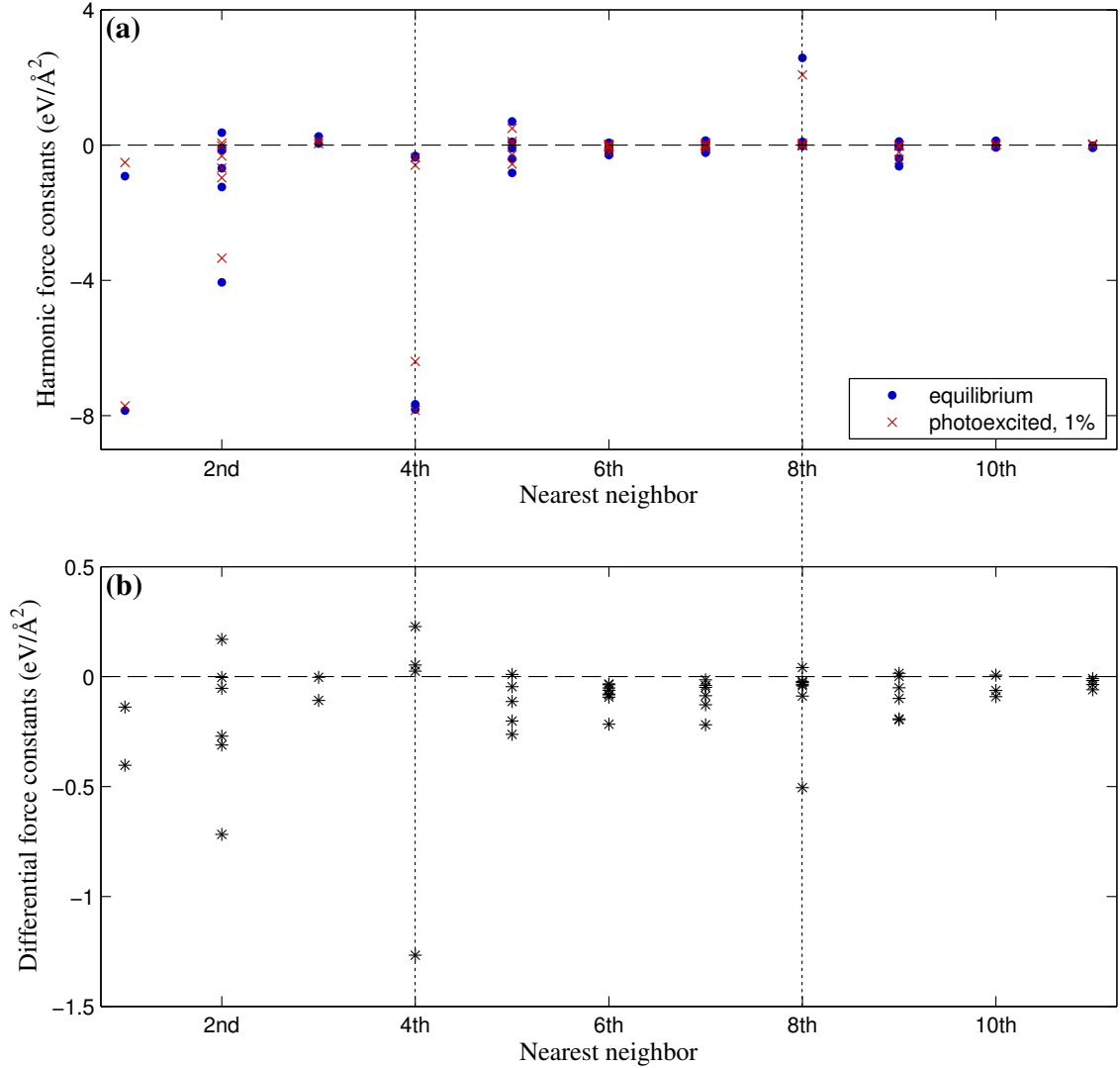

**Figure 3 | Comparison of calculated harmonic interatomic force constants for PbTe at equilibrium (using DFT) and under 1% photoexcitation (using CDFT). a,** Nonzero harmonic interatomic force constants (IFCs) calculated for both equilibrium (blue dots) and 1% photoexcited PbTe (red crosses) on 216-atom supercells. The IFCs for the first eleven nearest neighboring atoms are displayed. Particularly large IFCs are observed between 1st, 4th, and 8th nearest neighbors, which correspond to interactions along the <100> cubic direction responsible for PbTe's incipient ferroelectricity. The absolute values of these large IFCs notably decrease with photoexcitation, driving PbTe further away from the ferroelectric phase. **b,** Differences between the absolute values of the equilibrium and photoexcited IFCs in Fig. 3a. Notably large differences in the IFCs are seen for the 4th and 8th nearest neighbors.